\documentstyle[pra,psfig,aps]{revtex}
\begin{document}                
\newcommand{\ignore}[1]{} 
\def\a{$\alpha$}
\def\be{\begin{equation}}
\def\ee{\end{equation}}
\def\ba{\begin{eqnarray}}
\def\ea{\end{eqnarray}}
\preprint{DOE/ER/40561-16-INT98}
\title{Nonergodic Behavior of Interacting Bosons in 
Harmonic Traps}
\author{Thomas Papenbrock and George F. Bertsch}
\address{Institute for Nuclear Theory, Department of Physics, 
University of Washington, Seattle, WA 98195, USA}
\maketitle
\begin{abstract}
We study the time evolution of a system of interacting bosons in a 
harmonic trap. In the low--energy regime, the quantum system is not ergodic
and displays rather large fluctuations of the ground state occupation number.
In the high energy regime of classical physics we find nonergodic behavior
for modest numbers of trapped particles. We give two conditions that assure
the ergodic behavior of the quantum system even below the condensation 
temperature. 
\end{abstract}   
\section{Introduction}
The understanding of the formation and growth of atomic Bose Einstein 
condensates \cite{An,Brad,Davis,Rev} is of considerable interest and has been 
investigated experimentally and 
theoretically \cite{Stoof,Holland,Wu,Jaksch,Gardiner}. 
Bose Einstein condensates may be formed while the trapped bosons are in 
contact with a heat bath and a particle reservoir 
(e.g. via evaporative cooling \cite{evcool}) or from the 
evolution of a nonequilibrium state in an isolated system. 
In a recent experiment at the MIT \cite{mcMIT}, harmonically trapped bosons
were rapidly cooled below the transition temperature, and the subsequent 
relaxation towards a Bose Einstein condensate was observed and measured.
The existing theory uses the Boltzmann equation, which is fundamentally
based on the ergodic Sto{\ss}zahlansatz. However, this assumption deserves
closer scrutiny in the context of bosons in traps. In particular, the
motion of two particles confined to a harmonic trap is integrable 
for any interaction potential that depends only on the distance.
Closely related to ergodicity is the
question about the fluctuations of the ground state occupation number in an
isolated interacting system. For the non--interacting system, these 
fluctuations have been calculated recently \cite{Holthaus,Gajda}. 
Fluctuations of the interacting system are first computed here.

The paper is organized as follows. The low--energy quantum Hamiltonian and
observables of interest are introduced in the second section. In the third
section, we present numerical results of the time evolution of the ground 
state occupation number and its fluctuations for bosons in a harmonic and a
square well potential, respectively. 
The comparison with a chaotic and ergodic random 
matrix model is presented in section four. The structure of classical phase 
space for a system of harmonically trapped hard--sphere particles is discussed
in the fifth section. Finally, the results are summarized in the conclusion.  

\section{Hamiltonian and Observables}
In this section we describe the quantum Hamiltonian and the observables
that may indicate nonergodic behavior. 
We will consider interacting bosons confined to 
a harmonic potential or a square well potential. Magnetic traps, as
used in recent experiments, are very well approximated by harmonic potentials.
However, the square well potential is also of considerable
interest, since many theoretical results were obtained for such traps and since
the underlying classical system of hard--sphere bosons is chaotic and 
ergodic \cite{Sinai}.

Independent of the specific trap potential, the Hamiltonian may be written
\be
\label{ham}
\hat{H}=\hat{H}_0 + \hat{V}.
\ee
Here 
\be
\label{H0}
\hat{H}_0=\sum_j E_j \,\hat{a}_j^\dagger \hat{a}_j
\ee
is the one--body trap Hamiltonian and 
\be
\label{Hint}
\hat{V}=\lambda\sum_{i,j,k,l}V_{ijkl}\,\hat{a}_i^\dagger \hat{a}_j^\dagger 
\hat{a}_k \hat{a}_l.
\ee
the two--body interaction. The operators $\hat{a}_j$ and $\hat{a}_j^\dagger$ 
annihilate and create one boson in the single--particle trap state $|j\rangle$
with energy $E_j$, respectively. They fulfill the usual bosonic commutation 
rules. In what follows, the interaction is chosen to be a contact interaction.

We will study the quantum dynamics of a $N$--body system of bosons using
the occupation number representation. The basis of states is 
$|\alpha\rangle\equiv|n_0,n_1,\ldots,n_k\rangle$ with $\sum_{i=0,k}n_i=N$ and 
$\hat{a}_j^\dagger\hat{a}_j|n_0,n_1,\ldots,n_k\rangle
=n_j|n_0,n_1,\ldots,n_k\rangle$. Here $n_j$ denotes the occupation of the
$j^{\rm th}$ single particle state $|j\rangle$. Obviously, the trap Hamiltonian
is diagonal in this basis.

We also note that condensation occurs in $d$-dimensional harmonic traps 
for energies $E\approx N^{1+1/d}\hbar\omega$ \cite{Temp,Rev}, 
and this is the regime we are most interested in.
No Bose Einstein condensation occurs in square well potentials in two
spatial dimensions.

\subsection{Harmonic Trap}
To be specific, we now consider the case of an isotropic harmonic trap in
two spatial dimensions. The single particle states are  
$|j\rangle\equiv|n_j,m_j\rangle$ with energy $E_j=n_j\hbar\omega$ and 
angular momentum $m_j\hbar$ 
($m_j = -n_j, -n_j+2,\ldots,n_j-2,n_j$). We denote the oscillator wave 
functions as $\phi_j(\vec{x})\equiv \langle\vec{x}|j\rangle$.
The ground state energy is set to 
zero. In terms of the dimensionless coordinate $\vec{x}=\vec{r}/d_0$ (where 
$d_0\equiv\sqrt{\hbar/m\omega}$ sets the length scale of the trap), the contact
interaction (\ref{Hint}) has matrix elements
\ba
\label{matele}
V_{ijkl}&=&\int d^2x\,\phi_i^*(\vec{x})\phi_j^*(\vec{x}) \phi_k(\vec{x}) 
\phi_l(\vec{x})\nonumber\\
&=&\frac{\delta_{m_i+m_j}^{m_k+m_l}}{2}\int\limits_0^\infty
dx\,x^{-2}\,M_{\frac{n_i+1}{2},\frac{|m_i|}{2}}(x)\,
M_{\frac{n_j+1}{2},\frac{|m_j|}{2}}(x)\,\nonumber\\
&&M_{\frac{n_k+1}{2},\frac{|m_k|}{2}}(x)\,
M_{\frac{n_l+1}{2},\frac{|m_l|}{2}}(x).
\ea
Here $M_{\mu,\kappa}(x)$ denotes the 
Whittaker function \cite{Abra} and the $\delta$-function ensures 
conservation of angular momentum. The coupling is given
by $\lambda=4\pi \hbar\omega a/d_0$, where $a$ is the $s$--wave 
scattering length. Obviously, the total energy $E$ and the total angular 
momentum $M$ are conserved quantities. Furthermore, the motion of the 
center of mass decouples from the single--particle motion yielding a ladder
of states spaced by $\hbar\omega$ for each spatial dimension \cite{Fetter}.  
Pitaevskii and Rosch have shown, that 
Hamiltonian (\ref{ham}) possesses an additional $SO(2,1)$ 
symmetry \cite{Rosch97}. This leads to another
conserved quantity $B$ that can be associated with breathing modes. 
For fixed quantum numbers $M$ and $B$ the spectrum of Hamiltonian (\ref{ham})
consists of an infinite ladder with spacing $2\hbar\omega$. 
The contact 
interaction (\ref{Hint},\ref{matele}) is valid in a dilute system at low 
energies $\sqrt{\hbar\omega/E}\gg a/d_0$, i.e. the scattering length is small 
compared to the wavelength.

One may think of the above model
as a realization of a {\em three}--dimensional oblate trap with 
$\omega\ll\omega_z$ and $a/d_0\ll a\sqrt{m\omega_z/\hbar}\ll 1$. 
Under these conditions, the Hamiltonian~(\ref{ham}) provides a 
reliable approximation for a dilute gas of atoms 
confined in an oblate magnetic trap at sufficiently low excitation energy.
The trap used by the JILA group is closest to such a trap. 
 
We use in the numerical calculation a basis spanned by the eigenstates 
$|\alpha,E,M\rangle$ of the trap Hamiltonian~(\ref{ham}) 
with total energy $E\le E_{\rm max}$ and angular momentum $M$. 
($\alpha$ accounts for the
different states with quantum numbers $E,M$. Their number is $\Omega(E,M)$; 
in what follows we may suppress the dependence on $E,M$.)
Within this basis, the matrix of the interaction 
(\ref{Hint}) is set up as follows. After fixing the total 
angular momentum $M$ and the maximal energy $E_{\rm max}$ we choose one 
(arbitrary) $N$--boson state $|\alpha,E,M\rangle$ with $E\le E_{\rm max}$. 
We then act with the interaction 
(\ref{Hint}) onto $|\alpha\rangle$ and onto all the states created by 
this procedure until the space with energies $E\le  E_{\rm max}$ and fixed
angular momentum $M$ is exhausted. The resulting matrix is sparse.
As one check on the numerics we note that the obtained spectra
display ladders with spacings $2\hbar\omega$ or $\hbar\omega$ 
resulting from the breathing mode \cite{Rosch97} or the center of mass motion
\cite{Fetter}, respectively.   
Results obtained after the exact diagonalization of Hamiltonian (\ref{ham})
were found to be in good agreement with mean field theory even for rather
small systems \cite{Haugset}.

At fixed angular momentum, the spectrum of the trap (\ref{H0}) 
consists of equidistant shells of highly degenerate
states with energy $E=2K\hbar\omega$. 
The spacing between the shells is $2\hbar\omega$ (since even (odd) 
$E_{\rm max}$ requires even (odd) $M$). We label the different shells by
their energy quantum number $K$.
Increasing the coupling $\lambda$ shifts the entire spectrum 
towards higher energies and lifts the degeneracies 
partially. However, the spectrum still exhibits shell structure for not too 
large couplings and number of particles.

\subsection{Square Well Trap}

In case of a square well potential of width $d_0$, the creation operator 
$\hat{a}_j$ creates a boson in the single particle state 
$|j\rangle\equiv|n_{jx},n_{jy}\rangle$ with energy
$E_j=E_0(n_{jx}^2+n_{jy}^2)$. 
Here, $E_0=\hbar^2/2md_0^2$ sets the energy scale. $n_{jx}$ and $n_{jx}$
are the quantum numbers in the $x$ and $y$--direction, and the wave functions
are $\phi_j(x,y)=2\sin(n_{jx}\pi x)\,\sin(n_{jy}\pi y)$, 
where distances $0\le x,y\le 1$ are measured in units of $d_0$. 
States may be characterized by their behavior under reflection at the axes 
$x=1/2$, $y=1/2$. (We do not specify the behavior under 
reflection $x\leftrightarrow y$.)
The coupling is given by $\lambda=4\pi E_0 a/d_0$ in terms of the 
$s$--wave scattering length $a$.
In what follows we consider those states which have the same symmetry as the 
ground state. The Hamiltonian matrix is built up as for the harmonic trap. 
At zero coupling $\lambda$, 
the spectrum is highly degenerate and thus exhibits shell 
structure. With increasing coupling, most of the degeneracies are lifted, the 
entire spectrum is shifted towards higher energies, and the shells overlap 
very soon. This behavior is different from the harmonic trap where the shell 
structure persists even for relatively large values of the coupling. 

\subsection{Observables}

Nonergodic behavior manifests itself in the difference between the time average
of a dynamical observable and its ensemble average. The most interesting
observable is the ground state occupation number, 
$\hat{n}_0=\hat{a}^\dagger_0\,\hat{a}_0$. This observable has been measured, 
and its microcanonical fluctuations have been computed recently 
for the non--interacting system \cite{Holthaus}. 
The time evolution for its expectation value with the trap eigenstate state 
$|\alpha\rangle$ is 
\ba
\label{occ}
n_0^{(\alpha)}(t)&=&\langle\alpha|\hat{U}^\dagger(t)\,\hat{n}_0 \,\hat{U}(t)|
\alpha\rangle. 
\ea
Here 
\be
\hat{U}(t)=\exp\left[-i\hat{H}t/\hbar\right]
\ee
is the time evolution operator. In practice, we evaluate eq.~(\ref{occ}) by 
applying $U(\Delta t)$ with a suitable chosen $\Delta t$ onto
the initial state $|\alpha\rangle$ by using its Taylor expansion 
\footnote{We have used a step size $\lambda\omega\Delta t\approx O(1/10)$
to assure a fast convergence of the Taylor expansion.}. 
This procedure allows one to treat rather large matrices.
The unitarity of $\hat{U}(t)$ was numerically ensured by an error of the norm
$\langle\alpha \hat{U}^\dagger(t)|\hat{U}(t)\alpha\rangle$ that was smaller 
than one percent. The results are compared to the equilibrium value 
\be
\label{equi}
\overline{n_0}\equiv \frac{1}{\Omega}\sum_{\alpha=1}^{\Omega} 
\langle\alpha|\hat{n}_0|\alpha\rangle.
\ee
In the case of the harmonic trap potential, this equilibrium value is the 
ensemble average over all $\Omega(E,M)$ states within one energy shell. 
Note that the states within this ensemble have different values of the 
conserved quantity $B$ associated with the breathing mode.   
In the case of the square well potential the ensemble comprises all states
within an energy layer $\delta E\approx 5\pi^2\hbar^2/2md_0^2$ around
$E$ that have the same symmetry as the ground state. 
$\delta E$ is reasonably smaller than the ground state energy 
$2N\pi^2\hbar^2/2md_0^2$ even for modest numbers of particles. 
Note that these ensembles differ from the microcanonical ensembles which are
determined by energy only and do not depend on additional quantum numbers.
However, nonergodic behavior in a sector defined by additional quantum numbers
implies nonergodic behavior in the entire energy shell. 

A second and more quantitative indicator for nonergodic behavior is given 
by the fluctuations of the ground state occupation number. These are defined
by 
\be
\label{n0}
\delta n_0^2(\lambda)=\frac{1}{\Omega}
\sum_\Psi \langle\Psi|\hat{n}_0|\Psi\rangle^2
-\left(\frac{1}{\Omega}
\sum_\Psi \langle\Psi|\hat{n}_0|\Psi\rangle\right)^2,
\ee
where $|\Psi(E,M)\rangle$ are the eigenstates of the interacting system. 
Note that the fluctuations (\ref{n0}) depend on $\lambda$.

For an ergodic system, these fluctuations should be very small since
an expectation value should not depend on the particularly chosen state 
within an energy shell of $\Omega$ states. 
It is useful to compare the fluctuations for the interacting system with 
the fluctuations for the non--interacting system
\be
\label{deltan0}
\delta n_0^2(0)\equiv\overline{n_0^2}-\overline{n_0}^2=
\frac{1}{\Omega}
\sum_\alpha \left(\langle\alpha|\hat{n}_0|\alpha\rangle
-\overline{n_0}\right)^2.
\ee
To compute the fluctuations (\ref{n0}), we diagonalize the 
Hamiltonian (\ref{ham}) in the basis 
$|\alpha\rangle$ and obtain the eigenstates $|\Psi\rangle$. 

\section{Numerical Results}

We describe the numerical results obtained for the harmonic trap and the
square well potential.

\subsection{Harmonic Trap}

\begin{figure}
  \begin{center}
    \leavevmode
    \parbox{0.9\textwidth}
           {\psfig{file=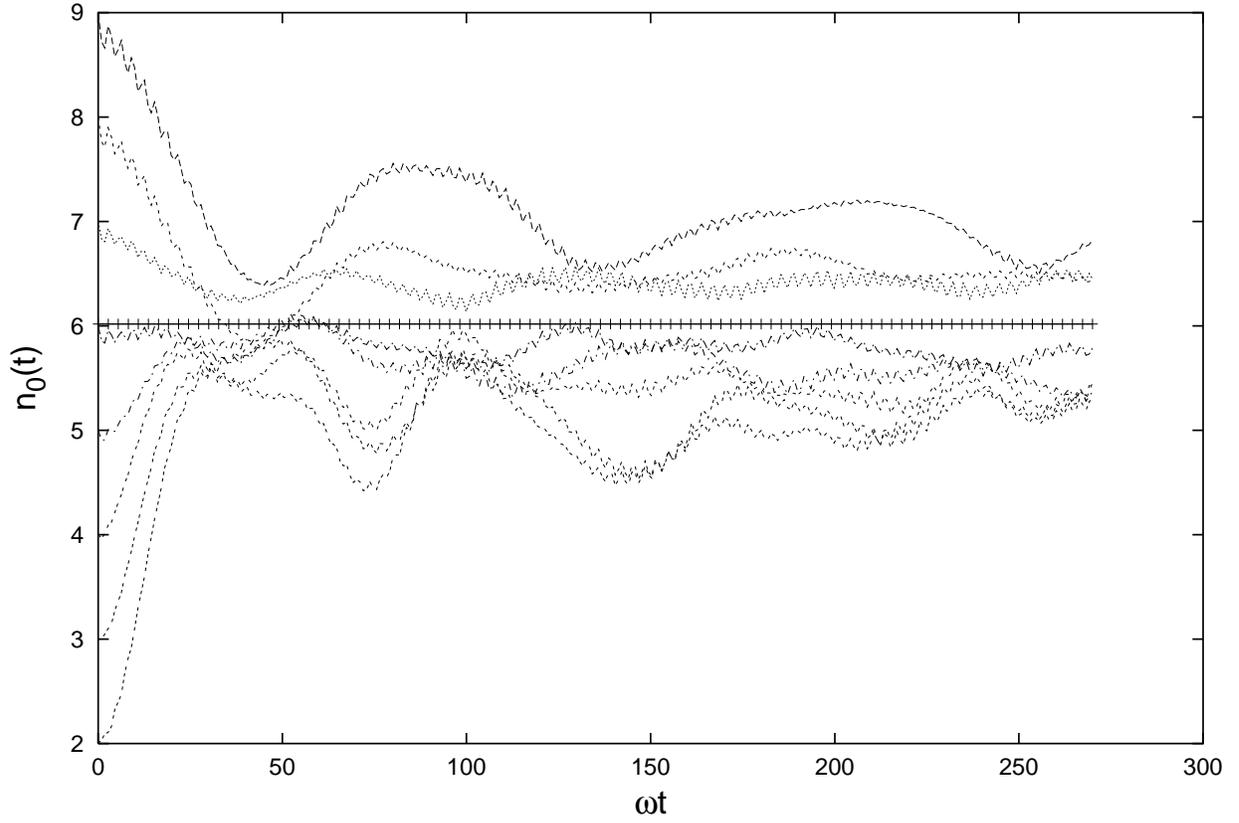,width=0.9\textwidth,angle=270}}
  \end{center}
\protect\caption{Time evolution (thin lines) of the ground state 
occupation number (\protect\ref{occ}) 
for different initial states of the $(K=4)$--shell 
that are initially away from the equilibrium value (thick line).
Results are obtained for a system of $N=10$ harmonically trapped bosons that
interact via $s$--wave scattering. The coupling is $\lambda=0.15\hbar\omega$.} 
\label{fig1}
\end{figure}

We choose $\lambda=0.15\hbar\omega$, $N=10$ and $E_{\rm max}=12\hbar\omega$. 
This space has a dimension of 1530 and according to the criterion given in 
Refs.~\cite{Temp} should exhibit condensation.  
Fig.~\ref{fig1} shows $n_0^{(\alpha)}(t)$ for a 
few states of the $(K=4)$--shell 
that have initially $2\le n_0^{(\alpha)}(t=0)\le N-1$. 
These states have occupation numbers that are initially 
different from the equilibrium value $\overline{n_0}$ defined in 
eq. (\ref{equi}). As shown in Fig.~\ref{fig1},
for $n_0^{(\alpha)}(0)<\overline{n_0}\quad$ 
[$n_0^{(\alpha)}(0)>\overline{n_0}]$ 
the ground state occupation number rises [falls] on a time scale involving 
many periods and saturates on a higher [lower] level. However, the 
saturation is lower [higher] 
than the equilibrium occupation number and is obviously different for 
different initial states. 
This indicates that the quantum system is not ergodic. 

Table.~\ref{tab1} shows the fluctuations (\ref{n0}) for different 
values of the coupling $\lambda$ and $N=10$ bosons. 
Again, we have chosen
$E_{\rm max}=12\hbar\omega$. With increasing $\lambda$ the spectrum is 
shifted towards higher energies but the shell structure persists.  
For nonzero $\lambda$ the fluctuations are 
suppressed in comparison to the non--interacting case, and are practically 
independent of $\lambda$ for $0<\lambda\le 0.2$. 
The last row of Table~\ref{tab1}
presents the fluctuations one would expect in a ergodic system. 
These findings confirm the results obtained from the time evolution.
They show that the quantum system is not ergodic in the regime of low energies
and modest number of bosons. 

We also consider the $N$--dependence of the fluctuations. Fixing 
$\lambda=0.025\hbar\omega$ and $E_{\rm max}=12\hbar\omega$, 
Table.~\ref{tab2} shows 
the ratio $\sqrt{\delta n_0^2(\lambda)/\delta n_0^2(0)}$ for different 
numbers of particles. 
One observes that the fluctuations increase with increasing $N$
and approach the fluctuations of the non--interacting system. This can be 
understood as follows. 
Since $E_{\rm max}$ has been fixed, the energy available per boson is
decreasing with increasing numbers of bosons and thus the regime $N\gg 1$
corresponds to the low temperature regime where most bosons are in the
condensate. The interactions induce mostly scattering within the condensate,
and this corresponds to the diagonal part of the interaction matrix 
(\ref{Hint}), which does not yield any change in the eigenfunctions.  

As shown in Section~\ref{SecClass} the classical system becomes chaotic for
$N^{5/2}(a/d_0)\sqrt{\hbar\omega/E}>1$. Thus, the classical system 
is chaotic for the values $N=40$, $\lambda=0.025\hbar\omega$ and 
$E=6\hbar\omega$
used in Table~\ref{tab2}. Nevertheless, this property of the classical system
is not reflected by the quantum system.  

\subsection{Square Well Potential}

\begin{figure}
  \begin{center}
    \leavevmode
    \parbox{0.9\textwidth}
           {\psfig{file=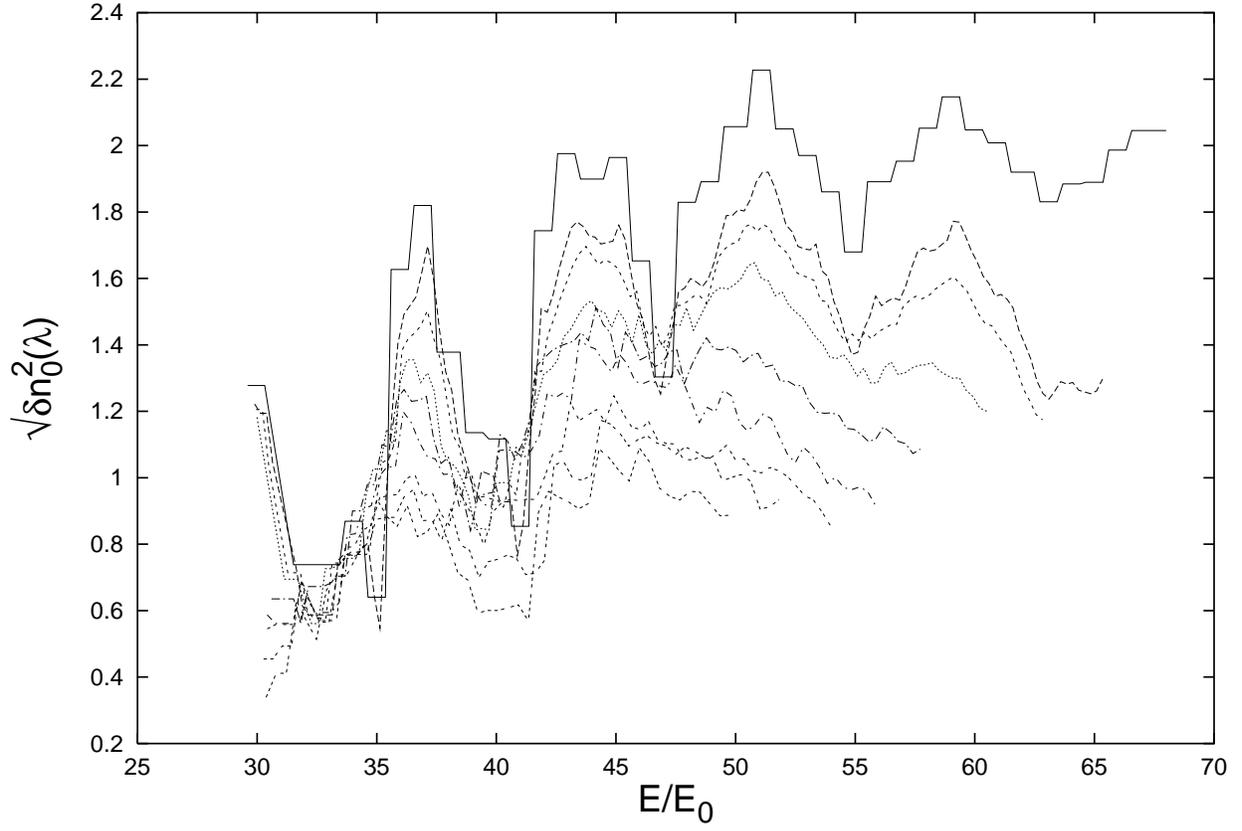,width=0.9\textwidth,angle=270}}
  \end{center}
\protect\caption{Fluctuations of the ground state occupation number 
$\sqrt{\delta n_0^2(\lambda)}$
for a system of $N=10$ bosons confined to a square well and different values of
the coupling $\lambda$. ($\lambda/E_0=0, {1\over 80}, {1\over 40}, 
{3\over 80}, {1\over 20}, {5\over 80}, {3\over 40}, {7\over 80}, 
{1\over 10}$ from top to bottom.) The energy is given in units of 
$E_0=\pi^2\hbar^2/2md_0^2$.}
\label{fig3}
\end{figure}

We next consider a square well potential, keeping the same number of bosons 
($N=10$) as before. Fig.~\ref{fig3} shows the fluctuations of the ground state 
occupation number as a function of energy for different values of the 
coupling $\lambda$. With increasing values of $\lambda$ the results have 
been shifted to lower energies, such that the average ground state occupation
number gets (almost) independent of $\lambda$. Fig.~\ref{fig3} shows that the 
fluctuations are relatively large even for the interacting system. 
Thus the quantum system is not ergodic at low energies. 
We recall that a classical system
of hard--sphere particles confined to a square well is chaotic and ergodic 
\cite{Sinai}. This classical behavior is expected to be followed
by the quantum system in the semiclassical limit 
\cite{LesHouches,GMW98} when the 
wave length is sufficiently small compared to other length scales including 
the scattering length. However, our contact interaction 
demands the opposite limit of long wave lengths.
One also recognizes in Fig.~\ref{fig3} that the fluctuations 
decrease with increasing $\lambda$. One may only speculate whether this 
behavior is tied to the ergodicity of the classical system.

Note that nonergodic behavior of a classically chaotic system is not
unexpected in the low--energy quantum regime. The same observation has been
made, e.g. for a system of one particle confined to the stadium billiard 
\cite{Casati}. 

\section{Random Interactions}
\label{rmt}
\begin{figure}
  \begin{center}
    \leavevmode
    \parbox{0.9\textwidth}
           {\psfig{file=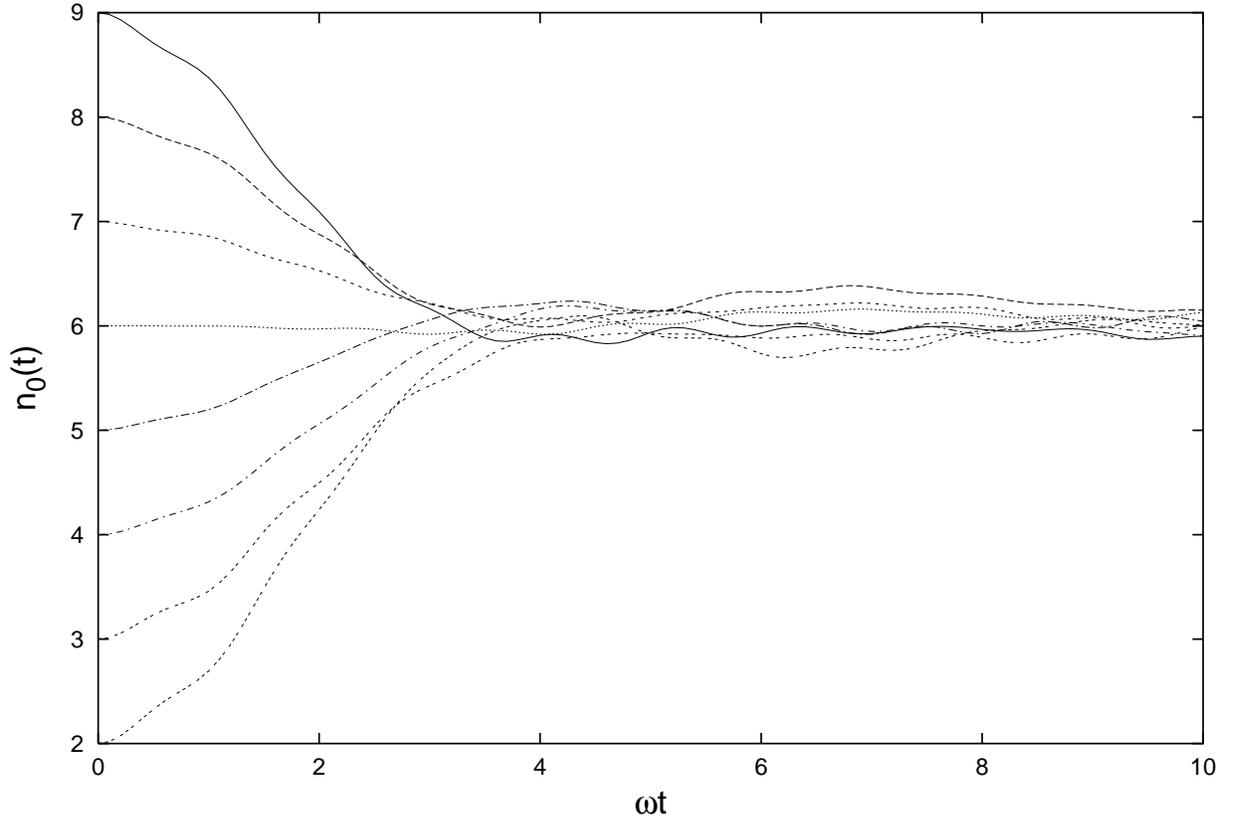,width=0.9\textwidth,angle=270}}
  \end{center}
\protect\caption{Time evolution (thin lines) of the ground state occupation 
number (\protect\ref{occ}) 
for different initial states of the $(K=4)$--shell 
that are initially away from the equilibrium value (thick line).
Results are obtained for a system of $N=10$ harmonically trapped bosons that
interact via random interactions.}
\label{fig2}
\end{figure}

We would now like to contrast the results presented in the last section 
with a quantum Hamiltonian that does display ergodic behavior. 
To this purpose, we restrict ourselves to the states within one energy shell 
and model the interaction by a ${\Omega(E,M)}$--dimensional random matrix 
which is drawn from the Gaussian Orthogonal Ensemble (GOE). Within one shell, 
the results depend only trivially on $\lambda$ and the variance 
of the random matrix elements. Our random matrix model is motivated
by the observation that, within the semiclassical regime, 
fluctuation properties concerning eigenvalues and wave functions of
classically chaotic systems are universal and coincide with those of the 
Gaussian random matrix ensembles \cite{Bohigas,GMW98}. Thus, one would 
expect that a 
random interaction would yield ergodic behavior. The appropriate time 
evolution of the ground state occupation number is 
shown in Fig.~\ref{fig2}. As expected, the equilibrium is practically reached 
by every initial state, and the fluctuations are relatively small. 

Using random matrix theory, we may also obtain analytical results for 
fluctuation of the ground state occupation number. 
Let $|\Psi\rangle=\sum_\alpha c_\alpha|\alpha\rangle$ be
an eigenvector of the random matrix Hamiltonian. Within the GOE 
the coefficients $c_\alpha$ are uniformly distributed over a 
${\Omega}$--dimensional sphere of unit radius. The normalized joint 
probability distribution for $k$ coefficients is given by \cite{Brody}
\be
P_k(c_1,\ldots,c_k)=\pi^{-k/2}\frac{\Gamma(\Omega/2)}{\Gamma((\Omega-k)/2)}
\left(1-\sum_{\alpha=1}^k c_\alpha^2\right)^\frac{\Omega-k-2}{2}.
\ee
To compute expectation values, the average over the energy shell is now 
replaced by the GOE average, and we denote the latter by brackets 
$\langle . \rangle$. This is justified for $\Omega\gg 1$ since the GOE becomes
ergodic in the limit of infinitely many levels, i.e. each of its members 
displays the same fluctuation properties as the ensemble \cite{GMW98}. 
The ground state occupation number of the eigenstate $|\Psi\rangle$ is 
\be
n_0^{(\Psi)}\equiv\langle\Psi|\hat{n}_0|\Psi\rangle=\sum_\alpha c_\alpha^2
n_0^{(\alpha)},
\ee 
and its GOE average is
\be
\langle n_0^{(\Psi)}\rangle=\int\limits_{-1}^1\,dc\,P_1(c)\,n_0^{(\Psi)}
=\overline{n_0}
\ee
as expected. Using $P_2$ we obtain for the variance
\be
\label{variance}
\delta n_0^2({\rm GOE})\equiv
\langle(n_0^{(\Psi)}-\langle n_0^{(\Psi)}\rangle)^2\rangle
=\frac{2}{\Omega+2}\delta n_0^2(0),
\ee
where $\delta n_0^2(0)$ is the variance of the non--interacting system given in
eq. (\ref{deltan0}). This shows that random interactions lead to a 
tremendous suppression in comparison to the non--interacting case and vanish
in the limit of infinitely many levels. Note that an expression similar to 
eq.(\ref{variance}) holds for any one--body observable that commutes with the
unperturbed Hamiltonian. 
Table~\ref{tab3} compares the fluctuations of the ground state
occupation number of the non--interacting system with the numerical results
obtained with a random Hamiltonian and with the analytical result 
(\ref{variance}). Obviously, the fluctuations of the non--interacting system
are much larger than those of the random Hamiltonian. One also recognizes
that the agreement between the analytical result and the numerical 
simulation improves with increasing dimension of the random matrix. 
These results show that the interaction mediated by $s$--wave scattering 
is very different from a chaos simulating random many--body interaction.

\section{Classical Phase Space}
\label{SecClass}
It would be interesting to investigate the ergodic properties also at higher
energies. Unfortunately, a treatment of the full many--body system is very
difficult outside the low--energy quantum regime. However, it is useful
to study the classical many--body system in more detail. Within the 
semiclassical regime, where the wave length is small compared to any 
length scale including the scattering length (i.e. 
$(a/d_0)\sqrt{E/\hbar\omega}\gg 1$ \footnote{This 
estimates is based on the assumption that one particle has the entire 
energy $E$. A more conservative estimate would use the fraction $E/N$ for
single--particle energies.}), the quantum system is expected to 
reflect the properties of the underlying classical system 
\cite{LesHouches,GMW98}. 
Sinai has shown that the classical system of hard--sphere particles 
in a square well potential is ergodic and chaotic \cite{Sinai}. We therefore 
consider the classical dynamics of a dilute system of hard--sphere particles 
of radius $a$ confined to a two--dimensional isotropic harmonic 
oscillator. At low energies, the corresponding quantum system 
is governed by the Hamiltonian (\ref{ham}).
Of course, in the low--energy limit, {\em any} short ranged two--body 
interaction with $s$--wave scattering 
length $a$ would yield the quantum Hamiltonian (\ref{ham}), and there is no
reason to choose a hard--sphere interaction. 
However, hard potentials induce more chaos 
than softer ones, and they are easier to treat. This justifies and 
motivates our specific choice. Nevertheless, most of the results derived 
below are valid for any two--body interaction that depends only on
the distance and vanishes for distances larger than the scattering length.

Let us first consider the two--body system with coordinates and momenta
$\vec{r}_i$ and $\vec{p}_i$  ($i=1,2$), respectively. 
Introducing coordinates and momenta  
$\vec{q}_\pm = (\vec{r}_1\pm\vec{r}_2)/\sqrt{2},\quad 
\vec{p}_\pm = (\vec{p}_1\pm\vec{p}_2)/\sqrt{2}$ shows that the the system is
equivalent to a system of two nonidentical and non--interacting particles 
in a harmonic oscillator
where the particle with coordinates $\vec{q}_-$ 'sees' a spherical symmetric 
hard core potential with radius $\sqrt{2}a$ in addition to the confining 
harmonic potential. In the new coordinate system,
single particle energies and single particle angular momenta are conserved, 
which renders the system integrable. 
Obviously, scattering occurs only if the angular
momentum of the particle with coordinates  $\vec{q}_-$ is sufficiently low, 
i.e. $l_-^2<4\hbar^2(a/d_0)^2[E/\hbar\omega-(a/d_0)^2]$. 
Let us assume that the initial
conditions of the two--body system are uniformly distributed over the 
shell of energy $E$. The probability of scattering is basically given 
by integrating $\Theta(4\hbar^2(a/d_0)^2[E/\hbar\omega-(a/d_0)^2]-l_-^2)$ 
over the energy shell in phase space. One obtains
\ba
\label{Prob}
P^{(2)}(\xi)=\left\{  
 \begin{array}{r@{\quad : \quad}l}
              1 & 1<\xi<2\\
 \frac{24/5}{(\xi-1)^{1/2}}-\frac{9}{\xi-1}+\frac{8}{(\xi-1)^{3/2}}
-\frac{3}{(\xi-1)^{2}}+\frac{1/5}{(\xi-1)^{3}}  & 2<\xi
 \end{array} \right.        
\ea       
where $\xi=(d_0/a)^2 E/(\hbar\omega)$ is a 
dimensionless parameter. Note that $\xi\gg1$ 
in the traps used in recent experiments. Obviously, the scattering probability 
is very small for two--body systems, and this is a consequence of the well 
known fact that the periods do not depend on the energy in harmonic 
potentials. For the $N$--body system we may use eq. (\ref{Prob}) to compute 
the probability of having at least one scattering. We give an estimate. 
On average, any two particles have a fraction $2/N$ of the total energy.
There are $N(N-1)/2$ different pairs of particles which may collide. Thus,
\be
\label{scatt}
P^{(N)}\approx \frac{N(N-1)}{2}\,P^{(2)}(2\xi/N).
\ee
A more accurate calculation of the leading term $\propto N^{5/2}\xi^{-1/2}$ 
in eq. (\ref{scatt}) confirms that the estimate is a good approximation.
Note that formulae eq. (\ref{scatt}) and eq. (\ref{Prob}) are valid for 
any short ranged two--body interaction that vanishes for distances larger 
than $a$. The absence of any scattering for some fraction of the 
energy shell is a consequence of the energy independence of the periods in 
the isotropic harmonic potential. Thus, the interacting classical system is 
not ergodic if the number of particles is not too large.

Let us also consider those regions of phase space that involve scattering 
among more than two particles. Starting trajectories in such regions, we 
computed the  Lyapunov exponent \cite{Benettin76} for systems with $2<N<25$ 
particles. As expected we found positive Lyapunov exponents. Thus, scattering
between more than two particles yields chaotic dynamics in the corresponding 
fraction of phase space. As a rule of thumb, eq.(\ref{scatt}) shows that 
the classical system becomes chaotic for 
$N^{5/2}(a/d_0)\sqrt{\hbar\omega/E}>1$.

Let us also discuss the case of a three--dimensional cylindrical symmetric 
trap with axial symmetry. In the absence of interactions the motion in the 
$z$--direction decouples from the motion in the radial plane. Since a 
scattering in three dimensions also is a scattering of the particles in 
the radial plane, the results (\ref{Prob},\ref{scatt}) derived in this 
section also hold for three--dimensional axially symmetric traps 
provided the total energy is replaced by the {\em radial} energy. 

Our results are consistent with the recent theoretical observation 
that the quasiparticle motion of collective and single--particle excitations 
of Bose Einstein condensates is only weakly chaotic \cite{Graham}. 

\section{Conclusion}

We have studied the ergodic properties of trapped bosons that interact 
via $s$--wave scattering. In the low--energy quantum 
regime we find nonergodic behavior for both the harmonic trap and the square
well potential. The nonergodic behavior may be seen in the difference
between time averages and ensemble averages and in rather large 
fluctuations of the ground state occupation number. For modest numbers of
harmonically trapped bosons the fluctuations are smaller than for the 
non--interacting system and almost independent of exact magnitude of the 
$s$--wave scattering length. For larger numbers of harmonically trapped bosons
the fluctuations of the interacting system approach the fluctuations of the 
non--interacting system in the low--temperature limit. This shows that 
even the many--body system is not ergodic at sufficiently low temperatures. 

The analysis of the classical phase space structure shows that, unlike 
square well potentials, harmonic potentials do not necessarily lead to chaotic
behavior of trapped interacting particles. Only for 
$N^{5/2}(a/d_0)\sqrt{\hbar\omega/E}>1$ a considerable fraction of classical 
phase space is chaotic. This is achieved with presently used traps 
if $N>40$ particles are trapped at temperatures of the order of the 
condensation temperature ($E\approx N^{1+1/d}\hbar\omega$). 
The quantum system is expected to follow the classical system in the 
semiclassical regime where the wave length is small compared to the scattering 
length. This is the case for $(a/d_0)\sqrt{E/\hbar\omega}\gg 1$, and 
$N\gg 3\cdot 10^4$ or $N\gg 10^4$ atoms have to be trapped in $d=3$ or 
$d=2$ dimensions, respectively.

The parameter that compares the strength of the interaction with the 
kinetic energy is $N a/d_0$ \cite{Rev}. For the experiments with rubidium 
at JILA \cite{An}, lithium at Rice University \cite{Brad}, 
and sodium at the MIT \cite{Davis}, one has
$N a/d_0 < 1, N a/d_0> 1$, and $N a/d_0\gg 1$, respectively. Our computations
are restricted to the weakly interacting regime $N a/d_0 < 1$, and as one 
consequence we do observe shell structure. 
It is not obvious, that this restriction is also
responsible for the nonergodic behavior observed in this work. The results
of Section~\ref{rmt} show that any small random interaction may render the
system ergodic since the noninteracting system is highly degenerate.  
Whether the quantum system is ergodic also in the low energy quantum regime 
for $N a/d_0> 1$ remains an open question. Given the fluctuations 
(\ref{variance}) for the system with random interactions on the one hand and 
the results \cite{Holthaus} for the non--interacting system on the other hand, 
experiments should be able to reveal how ergodic systems of trapped atoms 
really are.  

We have considered the fluctuation of the ground state occupation number
as one observable that is sensitive to nonergodic behavior. Chaotic systems 
exhibit level repulsion within sectors of definite symmetry and may 
be distinguished from integrable ones by their level 
statistics~\cite{Bohigas,GMW98}.
Due to its $SO(2,1)$ symmetry the two--dimensional harmonic trap 
with contact interaction is quite special. Sectors of 
definite symmetry (fixed angular momentum $M$ and fixed value 
of $B$) are ladders of levels with spacing $2\hbar\omega$ \cite{Rosch97} and 
are expected to display the same level statistics as a harmonic oscillator.

It is also interesting to discuss the growth of the condensate in 
harmonic traps. As long as 
the shells do not overlap and first order perturbation theory in $\lambda$ is 
valid, the quantum mechanical time scale set by the 
interaction is given by the product $1/\lambda\omega$. This may be 
compared to an approach using the Boltzmann equation, where the time scale
is proportional to $1/\lambda^2$ (when transition rates are obtained from
Fermi's golden rule). Thus, for sufficiently small values of 
$\lambda$, the use of the Boltzmann equation in combination with transition 
rates resulting from Fermi's golden rule is inappropriate. It yields
times for the condensate formation that are too large. This finding is a 
consequence  of the high degeneracy of the harmonic trap spectrum.  

\section*{Acknowledgement}

We acknowledge discussions with  Paulo Bedaque and Aurel Bulgac. 
This work was supported by the Department of Energy on 
Contract No. DE-FG-06-90ER-40561.

\begin{table}
\begin{tabular}{|l||c|c|c|c|}
                          & $K=2$   & $K=3$  & $K=4$  & $K=5$  \\\hline\hline
$\lambda=0.0\hbar\omega$  &   0.83  &  1.11  &  1.32  &  1.51  \\\hline
$\lambda=0.025\hbar\omega$&   0.56  &  0.55  &  0.59  &  0.64  \\\hline
$\lambda=0.05\hbar\omega$ &   0.54  &  0.55  &  0.60  &  0.64  \\\hline
$\lambda=0.075\hbar\omega$&   0.54  &  0.55  &  0.60  &  0.65  \\\hline
$\lambda=0.1\hbar\omega $ &   0.53  &  0.55  &  0.59  &  0.64  \\\hline
$\lambda=0.15\hbar\omega$ &   0.53  &  0.55  &  0.59  &  0.63  \\\hline
$\lambda=0.2\hbar\omega$  &   0.52  &  0.55  &  0.58  &  0.64  \\\hline
GOE                       &   0.35  &  0.27  &  0.18  &  0.12  \\
\end{tabular}
\protect\caption{Fluctuations of the ground state occupation number 
$\sqrt{\delta n_0^2(\lambda)}$ for different shells (labeled by $K$) 
and different values of the coupling $\lambda$. 
Systems of $N=10$ bosons are considered. 
For $\lambda=0$, the states of the $K$th 
shell have the energy $E=2K\hbar\omega$. The shells comprise
$\Omega(E=2K\hbar\omega,M=0)=9, 31, 109, 339$ states for $K=2, 3, 4, 5$, 
respectively. The last row shows results expected for an ergodic system.} 
\label{tab1}
\end{table}

\begin{table}
\begin{tabular}{|c||c|c|c|c|}
        & $K=2$  & $K=3$  & $K=4$  & $K=5$  \\\hline\hline
$N=12$  &  0.71  &  0.56  &  0.51  &  0.46  \\\hline
$N=15$  &  0.77  &  0.65  &  0.58  &  0.54  \\\hline
$N=20$  &  0.83  &  0.73  &  0.68  &  0.63  \\\hline
$N=25$  &  0.86  &  0.78  &  0.73  &  0.69  \\\hline
$N=40$  &  0.91  &  0.85  &        &        \\
\end{tabular}
\protect\caption{Fluctuations of the ground state occupation number 
$\sqrt{\delta n_0^2(\lambda)}$
normalized by the fluctuations of the non--interacting system 
$\sqrt{\delta n_0^2(0)}$ for different shells (labeled by $K$) and 
different numbers $N$ of bosons. The coupling is fixed to 
$\lambda=0.025\hbar\omega$.
The shells comprise $\Omega(E=2K\hbar\omega,M=0)=9, 31, 109, 339$ states for 
$K=2, 3, 4, 5$, respectively.} 
\label{tab2}
\end{table}

\begin{table}
\begin{tabular}{|c||c|c|c|}
                & $K=4$      & $K=5$   & $K=6$    \\\hline\hline
HO              & 1.32       & 1.51    & 1.64     \\\hline 
RM              & 0.20       & 0.112   & 0.072    \\\hline 
GOE             & 0.18       & 0.116   & 0.072    \\
\end{tabular}
\protect\caption{Fluctuations of the ground state occupation number 
in different shells labeled by $K$. Results are shown for the harmonic 
oscillator (HO), harmonic oscillator with random interaction (RM), and 
the analytical result (\protect\ref{variance}) derived from random matrix 
theory (GOE). The shells comprise $\Omega=109,339,1039$ states for $K=4,5,6$, 
respectively.}
\label{tab3}
\end{table}

\end{document}